\documentclass [12pt] {article}

\begin {document}

\noindent {\Large \bf Complex energy flows in non-empty space for double unification of particles with fields and charges with masses}

\bigskip
{I.E. Bulyzhenkov}
\smallskip

\noindent{{\small Moscow Institute of Physics and Technology, \\9 Institutskiy per., 
 Dolgoprudny, Moscow reg., 141700, Russia; \\ Lebedev Physics Institute RAS, 53 Lininsky pros., Moscow, 119991, Russia
         }}
\bigskip

{\small {{\bf Abstract}. Non-empty space reading of Maxwell equations as local energy identities explains why a Coulomb field is carried rigidly by electrons in experiments. The analytical solution of the Poisson equation defines the sharp radial shape of charged elementary densities which are proportional to  continuous densities of  electric self-energy. 
Inward and outward longitudinal waves within the continuous electron reshape its radial energy structure in external fields.
 Both Coulomb field and radial charge densities are free from energy divergences. Non-empty space of electrically charged mass-energy  can be described by complex analytical densities resulting in real values for volume mass integrals and in imaginary values for volume charge integrals. Imaginary electric charges in the Newton gravitational law comply with real Coulomb forces. Unification of forces through complex charges rids them of  radiation self-acceleration.
\bigskip

}}



\section{Introduction}

Recent measurements of dynamical electric fields associated with moving  charges \lq\lq{}support the idea of a Coulomb field carried {\it rigidly} by the electron beam\rq\rq{} \cite {Piz}. This experimental result seems  natural only in non-empty space physics, where charged radial fields represent distributions of continuous elementary matter in the very radial structure of its Coulomb/Newton fields. Non-empty space (inferred first by Aristotle as a continuous material plenum beyond observations) can be better justified through extended electric charges in the laboratory, rather than through extended masses in observed weak gravitation. Since the days of the ancient Greeks, non-empty space for physical reality had been  repeatedly claimed by many thinkers, including Einstein: \lq\lq{}A coherent field theory requires that all elements be continuous... And from this requirement arises the fact that the material particle has no place as a basic concept in a field theory. Thus, even apart from the fact that it does not include gravitation, Maxwell's theory cannot be considered as a complete theory\rq\rq{} (translation \cite{Ton}). In this paper we try to comment quantitatively on this Einstein\rq{}s vision of a coherent theory of pure classical fields which could indeed strike out the particle as an independent concept.  

\section {Fields in Maxwell-Lorentz equations do \\ define radial shapes of  charges}
The postulated  point-particle paradigm results not only in the mathematical divergence of the Coulomb energy, but also in the physical inconsistence of the microscopic electron theory. Any point source in the microscopic Maxwell-Lorentz Equations may be considered as \lq\lq{}an attempt which we have called intellectually unsatisfying\rq\rq{} according to De Broglie \cite{Bro}. Einstein also criticized his 1915 field equation for the point gravitational source: \lq\lq{}it resembles a building with one wing built of resplendent marble and the other built of cheap wood\rq\rq{} (tran\-slation \cite{Ton}). 
 The Dirac delta-operator formalism for the point charge seems a provisional modeling of physical reality until local analytical charge-field relations can be finally proposed for  Maxwell's electro\-dynamics.

The continuously distributed elementary charge was reasonably inferred by Mie in order to derive properties of charges from properties of fields (and potentials) and to overcome the energy divergence flaw  in the Coulomb field center. Regretfully, 
the \lq{}Theory der Matter\rq{} \cite {Mie} had not found the appropriate (logarithmic) post-Coulomb potential in 1912-1913.  And the  promising non-empty space concept had not been justified before the mathematical era of quantum non-locality. The formal probabilities for  the delta-operator \lq{}dice\rq{} in the same  empty space arena postponed the search of rigorous analytical solutions for Mie\rq{}s nonlocal matter. In 1984 Schwinger intuitively proposed  \cite {Sch} to extend the point classical electron by the exponential radial distribution like in the Yukawa field potential. However, the annoying problem of the unphysical point source and the Coulomb energy divergence are still unresol\-ved satisfactorily within the trustful mathematical formalism of classical fields.  These permanent challen\-ges of contemporary theoretical and mathema\-tical physics may be considered as a motivation for our reinforcement of Mie and Schwinger nonlocal (astro)particle by the continuous radial density over the infinite world volume or the entire Universe.

Let us revisit classical electromagnetic equations regarding coherent solu\-tions for a distributed particle and its field. By following Mie, we assume that it is possible to relate analytically  electric, ${\vec d}(x)$, and magnetic, ${\vec b}(x)$, field intensities in the Maxwell-Lorentz equations \cite{Lor}, 
\begin {eqnarray}
\begin {cases} { 
 div\ {\vec {d}}({{\bf x}},t) = 4\pi\rho({{\bf x}},t) \cr
div\ {\vec b}({{\bf x}},t) = 0 \cr
c\ curl\  {\vec b}({{\bf x}},t)  = 4\pi\rho ({{\bf x}},t){\vec v} + \partial_t  {\vec {d}}({{\bf x}},t)\cr
c\ curl\ {\vec {d}} ({{\bf x}},t)  = - \partial_t {\vec b}({{\bf x}},t),\cr}
\end {cases}
\end {eqnarray}
 to the local charge, $\rho({{\bf x}},t)$, and current, $\rho({{\bf x}},t){\vec v }$, densities of the extended electron. Contrary to Lorentz, the electron's charge distribution  was not postulated by Mie within a microscopic spatial volume (electron's fields were assigned by Lorentz to charge-free regions or to the supposed empty space around charges). The mathematical equation $div \ {\vec {d}}({{\bf x}},t) = 4\pi\rho({{\bf x}},t)$ can be rigorously resolved under the non-empty space assumption by admitting that the elementary charge and its field coexist together in all space points ${\bf x}$ of the infinite Universe. In other words, we tend to maintain the Mie (and Einstein) idea that the elementary electric (and gravitational) charge is to be integrated into its spatial field structure with instantaneous local relations between scalar functions $\rho({{\bf x}},t) $ and ${\vec { d}}({\bf x},t)\cdot {\vec {d}}({\bf x},t). $ Then locally paired, {\it rigid} motion of Coulomb radial fields and electron\rq{}s radial shapes can    
maintain  the precision experiment \cite{Piz} as a direct justification of the non-empty space paradigm for physical reality. 

The mass density $m_o{n}(r)$ of the distributed radial electron with the analytical density ${n}(r)$ possess the same active mass-energy density as the passive mass-energy of electron's gravitational field (due to the Principle of Equivalence). Similarly, the charge density $\rho(r) = q{ n}(r)$ of the same electron should possess a self-energy density which is to be equal to the Coulomb field energy density. Therefore,  the passive charge density of the extended particle  
 is to be proportional to the active field energy density
\begin {equation}
 \rho({{\bf x}},t)=  
 [\pm \vec{d}({{\bf x}},t)]^2/4\pi\varphi_o
\end {equation}
 in the rest frame of reference. Such physical relations between Coulomb\rq{}s  field intensity and particle\rq{}s density can be expected among correct solutions to Maxwell\rq{}s equations if the Einstein Principle of Equivalence works for electricity. Moreover, the electric charge $q$, with self-energy $q\varphi_o$, and its density $\rho$, with self-energy density $ \rho\varphi_o$, has the energy nature under the universal potential $\varphi_o = const $ like a mass has the energy nature due to $E= mc^2$.  Maxwell-Lorentz's equations  can  be equally discussed as for electric currents with the electric density $ \rho$ as well as for energy flows with the electric energy density $ \rho\varphi_o$.  In general, electric self-energy $E_q= q\varphi_o$ of the charge $q$ should be added to the  mechanical self-energy $E_m = mc^2$ of a charged body. 

One can say that EM energy flows with their natural conservations/dissi\-pations are more fundamental for unification of CED and thermodynamics than electric currents with the well defined conservation of charges but ill-defined dissipation notions. Anyway, CED  can fluently employ the charge self-energy (justified so far only in QED) in Mie's non-empty space with material fields. Recall, that these fields can represent continuous particle-charges. The delta-operator description of localized char\-ges, say on separated capacitor's plates, can also introduce the electrostatic energy of such point charges in addition to the equal amount of their Coulomb field energy (separated spatially from such point charges). Contrary to diverging self-energies of point particles, self-energy of the spatially distributed charge can be finite and can be described through analytical densities with clear physical meaning.

The local equality of the charge energy density, $\rho \varphi_o$,  and the Coulomb field energy density, ${\vec d}^2/4\pi $, assumes finite electrostatic energy integrals of an elementary continuous carrier of electricity,  
\begin {equation}
\int\!\!  \frac { {\vec {d}}^2 ({{\bf x}},t)}{4\pi}dv = \varphi_o \!\! \int\!\! \rho ({{\bf x}},t)dv \!\equiv \varphi_o q \neq \infty.
\end {equation}      

One can use $\int\!\! \rho ({{\bf x}},t)dv \equiv q \equiv ie \Rightarrow - ie_o$ 
for the \lq{}negative\rq{} elementary charge of the electron. Its self-potential $\varphi_o \equiv  E_q/ q = q/ r_o = c^2/{\sqrt G}$
 we defined via the light speed limit $c$ and the Cavendish constant $G$. The similar charge to spatial scale ratio, ${\sqrt G}m/r_m = c^2/{\sqrt G}$, arises in  gravitation of continuous masses \cite {Bu3}, where  $mc^2 \equiv {\sqrt G}m \varphi_o $ is the body relativistic energy. 
We keep the fundamental potential $\varphi_o \equiv c^2/{\sqrt G} = 3.48 \times 10^{27} StatV = 1.04 \times 10^{27} V$ for self-energies of both mechanical and electric charges,
\begin {equation}
E = ({\sqrt G}m + q)\varphi_o = mc^2 + ieG^{-1/2}c^2,
\end {equation}
in order to extend the Einstein mechanical formula on electricity.  

Despite the electron possesses  in (4) only imaginary electric energy ($ - i\cdot1,04 \times 10^{24} KeV$ of the imaginary charge $q =-ie_o$) next to 511 $KeV$ of the real mechanical energy, paired interactions of imaginary electric charges correspond to real Coulomb forces and to real  interaction energies. Contrary to an electric charge defined in real numbers, like the inertial mass, the imaginary electric charge exhibits a correct direction of the radiation self-force, which is proportional to $q^2 = -e^2_o < 0$. Ultimately, joint densities of gravito-mechanic and electric charges can be described in complex functions. Real integrals over spatial densities provide masses of energy carriers, imagi\-nary integrals -  electric charges.  

\section {Inward and outward longitudinal waves \\ within the continuous electron}

Sphere radial currents or moving electric densities under strict spherical symmetry do not generate magnetic fields in Maxwell\rq{}s electrodynamics as is well known. Such magnetic field free currents  may equally have inward and outward directions with respect to a center of the radial electron. Time-varying currents of charged densities should be balanced by time-varying Maxwell displacement currents. In other words, the static radial shape of the elementary charge (associated with its static Coulomb field)  may be accompanied by radial wave modulations with longitudinal electric induction.       

Periodical inward, $(-{\hat {\bf r}}V_{in})\rho_{in} ({{r}},t)$, and outward, ${\hat {\bf r}}V_{out}\rho_{out} ({{ r}},t)$, currents, where ${\hat {\bf r}}\equiv {\vec r}/r$, are balanced by radial displacement currents $\partial_t  {{d_{in}}}({{r}},t){\hat {\bf r}}$ and  $\partial_t  {{d_{out}}}({{r}},t){\hat {\bf r}}$, respectively, with $div [{{d_{in}}}({{r}},t){\hat {\bf r}}] = 4\pi\rho_{in} ({{r}},t) $ and 
$div [{{d_{out}}}({{r}},t){\hat {\bf r}}] $ $= 4\pi\rho_{out} ({{r}},t)$. Electri\-cally charged radial densities with zero velocities and zero magnetic fields from symmet\-rical spherical currents do not contribute to the Maxwell equation for radial modulations of the continuous electron:  

\begin {eqnarray}
0 = 4\pi [\rho_{out} ({{r}},t){{\bf V}_{out}}+\rho_{in} ({{r}},t){{\bf V}_{in}}]+ 
\partial_t  [{{d_{out}}}({{r}},t){\hat {\bf r}}  +   {{d_{in}}}({{r}},t){\hat {\bf r}}]\cr =   \{|{\bf V}|\partial_r r^2[{{d_{out}}}({{r}},t)-{{d_{in}}}({{r}},t)] + r^2\partial_t  [{{d_{out}}}
({{r}},t) + {{d_{in}}}({{r}},t)]\}{\hat {\bf r}}/ r^2. \cr
\end {eqnarray}

This classical equation for balanced radial currents with the spherical symmetry has  wave solutions along both outward and inward directions,
\begin {equation}
\begin {cases} { {\vec d}_{out}(r,t) =   {\hat {\bf r}}q a_{out}   cos (\omega t - {k} {r} + \varphi_{out})/{r^2 } \cr
{\vec d}_{in}({{r}},t) = -  {\hat {\bf r}}q a_{in} cos  (\omega t + {k} {r} + \varphi_{in})/{r^2 }\cr}
\end {cases}
\end {equation}
These non-empty space modulations with  $k = \omega / C_l $  and the longitudinal wave speed $C_l = |{\bf V}_{out}| = |{\bf V}_{in}|$ are responsible for dynamical reshaping of  the continuous elementary charge $q$ in external fields. Such longitudinal waves within the elementary continuum keep its energy, while transverse electromagnetic waves are responsible for energy exchanges between different carriers of energy.  

In general, the dimensionless  wave amplitudes $a_{out}$ and $a_{in}$ may have different complex values related to power of longitudinal waves within the extended electron. However, there are no internal power flows within the elementary continuum in equilibrium. In this case, superposition of outward and inward waves (6) can exist in a form of  the following standing wave, 
\begin {equation}
{\vec d}_{stan}(r,t) =  \frac {q a{\hat {\bf r}}}{r^2}   [cos (\omega t - {k} {r}) - cos (\omega t + {k} {r}) ] =
\frac {2qa {\hat {\bf r}}}{r^2} sin kr\cdot  sin \omega t.
\end {equation} 

Now one can compute time-averaged energy of one standing wave modula\-tion within  the elementary charge volume,
 \begin {eqnarray}
 E_{a,\omega} =\int_o^\infty  \frac {(2qa)^2  sin^2 kr <sin^2 \omega t>_t  }{4\pi r^4}  4\pi r^2 dr  \cr =  2q^2a^2\int_o^\infty dr\frac {sin^2 kr}{r^2} = \pi a^2 q^2 \frac {\omega}{C_l}
\end {eqnarray} 
This energy is proportional to the frequency $\omega = C_l k$ of the wave excitation. A set of internal standing waves (7) with different frequencies $\omega$ and ampli\-tudes $a$ may play in Maxwell\rq{}s electrodynamics the same role as the foam of virtual photons next to the quantum electron in wave mechanics. De Broglie was first who assigned the internal wave process to each elementary particle with inertia in 1923. And later he specified this wave process for his unique physical interpretation of quantum phenomena \cite {De}. We tend to support the internal wave process [8] in therms of radial wave modulations (6)-(8) which shed some light on the celebrated double-slit experiment. In fact, not only quantum mechanics but also Maxwell\rq{}s theory of classical electric fields and electric energy flows may comment on the wave motion of continuous electrons. One day classical and quantum electrodynamics will lead to similar wave physics for the elementary charge. 

Transmission of energy-momentum from internal (virtual) longitudinal waves (6) to free transverse photons (or to other carriers of energy-momen\-tum) under proper interactions could reveal not only the well-established retarded wave signal from the particle center of spherical symmetry, but also a so-called advanced wave due to inward pulsations toward this center of  the extended charge.
By taking conventionally $C_l \Rightarrow c = 3\times 10^8 m/s$ for material space of the radial particle, one get guidelines to construct experimental facilities to probe three responses from electron or hadron beams: retarded (wave), instantaneous (Coulomb), and advanced (wave).  Anyway, the numerical value $C_l$ for speed of internal longitudinal modulations of extended charges is a challenge for contemporary electrodynamics.

\section {Logarithmic potential rids of Coulomb \\ energy divergence}

The radial  post-Newton potential $W({\bf x}) = - \varphi_o ln [1 +{\sqrt G}M/ |{\bf x} | \varphi_o]  \neq const $ of the active radial mass-energy  could be introduced \cite {Bu3} for interaction with other (passive) radial charges ${\sqrt G} m$. The mechanical self-energy $\varphi_o {\sqrt G} m = mc^2$ of the charge ${\sqrt G} m$ arises to compensate  the integral self-action of these charge densities $\rho_m ({\bf x}) = m r_o/4\pi {\bf x}^2 (r_o + |{\bf x}|)^2 $ in its interaction potential $W({\bf x})$,
\begin {equation}
\varphi_o {\sqrt G} m \equiv    \int \varphi_o \rho_m ({\bf x}) d^3x = -   \int  W ({\bf x}) \rho_m ({\bf x}) d^3x . 
\end {equation}
 This universal compensation of  Einstein\rq{}s mechanical self-energy $mc^2$ by the self-interaction of mass densities in their gravitational potentials, $\int [\varphi_o + W ({\bf x})]\rho_m d^3x \equiv 0$, takes place due to the integral mathematical equality $\int_o^\infty dx [ln (1 + x^{-1})]/(1+x)^2$ $\equiv \int_o^\infty dx/(1+x)^2$ $ = 1$.
Similarly, radial electric densities in their strong-field logarithmic potential $W ({\bf x})$ should  also lead through self-interactions to finite self-energy of the elementary continuous charge rather than to Coulomb energy divergence of the point charge. 

The divergence-free distributions (2)  for a steady continuous energy carrier  \lq{}at rest\rq{} (${\vec v}= {\vec b} = 0$) complies with  $div  \ {\vec {d}}(r) = 4\pi \rho(r)$ under the radial solution for field and charge densities, 
\begin {eqnarray}
\begin {cases} {
{\vec {d}}(r) =  {q {\hat {\bf r}} } / {r(r+r_o)},  \cr 
\rho (r)  =  {q r_o } / {4\pi r^2(r + r_o)^2}, \cr }
\end {cases}
\end  {eqnarray}
where the  particle's density $ n(r) = {r_o } / {4\pi r^2(r + r_o)^2}$ in non-empty space  replicates  
the Dirac operator density $\delta(r) $ under the empty space approxima\-tion of reality.  The exact solution (10) matches the finite balance of electric energy (3) and rids Maxwell electrodynamics of the Coulomb energy diver\-gence.

 The field flux of the radial charge distribution $\rho (r)$ depends on a selected radius R for a Gaussian sphere,
 \begin {eqnarray}
 4\pi R^2 {\hat {\bf r }}{\vec d}(R)\equiv {{q(R)}}= \int_0^R {{\rho(r)}}4\pi r^2dr \cr = \int_0^R{{q(\infty)r_odr}\over (r+r_o)^2}=q(\infty){R\over R+r_o}, 
    \end {eqnarray} where the complex number $r_o$ defines the half-charge scale $|r_o|$ for the infinite spatial (astro)distribution (10) of the  total charge $q \equiv q(\infty)$. In other words,  any non-zero value  $r_o  \neq 0$ unavoidably results in the global astrodistribution of the elementary charge over the entire Universe. Therefore, the non-point Maxwell electron (10) cannot be localized in principle  within the microscopic (and even  macroscopic) volume assumed in the Lorentz model. The classical charge density is rigidly bound with the electric field energy density, $\rho \propto {\vec {\rm d}}^2$. The equivalent charge and its field  counter-flows of imaginary electric energies fill non-empty world space  continuously together with collinear (residual) co-flows of real mass-energies within the elementary radial carrier \cite {Bu3}. The exact mathematical solution (10) to the classical field equations (1) admits the global material overlap of all elementary charges and rejects the empty space simplification of reality through separated fields and particles. 
  
The gravity-type Poisson equation,  $\nabla^2 W = 4\pi {\rho}   \Rightarrow \varphi_o^{-1} (\nabla W)^2  $, reads the imaginary (astro)electron as a non-linear field composition with respect to the radial field intensity ${d}(r)$ or the electron\rq{}s interaction potential $W(r)$,
 with ${d}(r) = -\partial_r W (r)$ and 
\begin {eqnarray}
\frac { \partial_r [r^2 \partial_r W(r)] }{4\pi r^2}=  \frac { [\partial_r W (r)]^2} {4\pi \varphi_o}=  {{q r_o}\over \  4\pi r^2 (r + r_o)^2}.  
\end {eqnarray}
 This non-linear equation reveals the imaginary post-Coulomb potential with the negative,  gravitational sign for the continuous radial carrier of electricity,
\begin {equation}
W (r) = - {  q \over  r_o} ln  \left (1 + {r_o\over r} \right ) \equiv - {\varphi_o } ln  \left (1 + \frac {q}{ \varphi_o r} \right ).
  \end {equation}

This logarithmic potential reproduces the regular Coulomb law on measu\-rable scales with an opposite, Newtonian sign  $(- q/r_o) ln [(r + r_o)/r] $ $\approx$ $ (-q/r) $, when $r_o \rightarrow 0$. The Newton-type potential (13) leads 
to real Coulomb forces, 
  $q_1 [-\nabla (-q_2r^{-1}_o)ln (1+ r_or^{-1})] =  ie_1(-ie_2){\hat {\bf r}}/ r(r+r_o)$ 
$\approx e_1e_2{\hat {\bf r}}/ r^2$, with mutual repulsion of like electric charges $e_1$ and $e_2$, and attraction of unlike ones.  The radiation self-force, $f_{rad} = 2q^2{\ddot {\bf v}}/3c^3 =  2(-ie_o)^2{\ddot {\bf v}}/3c^3 =- 2e_o^2{\ddot {\bf v}}/3c^3$, takes the physical, damping direction for the imaginary charge, contrary to a mathematical self-acceleration of the real electric charge.

The gravitational sign in the Poisson equation for imaginary electric energies suggests to inverse positive and negative imaginary densities of  electric charges, $\rho \rightarrow -\rho$, in the Maxwell-Lorentz equations (1), which are irrelevant to the conventional signs of electrons and protons. 
The electrostatic solution for charged imaginary densities in (1) with  the gravitational type direction of the radial Coulomb field, ${\vec {D}}({\bf x}) =  - {Q {\hat {\bf x}} } / {|{\bf x}|(|{\bf x}|+Q \varphi_o^{-1})}$,  facilitates the unification of gravity and electricity on the basis of one  complex energy charge $Q\varphi_o \equiv (q_m+iq_e)c^2/{\sqrt G} \Rightarrow ({\sqrt G} m + ie)c^2/{\sqrt G}$. The Maxwell-Lorentz equations (1) for the massless  elementary charge can be extended on the complex energy flows of the radial carrier  with the elementary charge density $\rho ({{\bf r}},t) \equiv ({\sqrt G} m + ie)n ({{\bf x}},t)= - div\ {\vec {D}}({{\bf x}},t)/ 4\pi$,
\begin {eqnarray}
\begin {cases} {
  div\ {\vec B}({{\bf x}},t)\varphi_o = 0 \cr
 c\ curl\  {\vec B}({{\bf x}},t)\varphi_o  = {\vec v}div\ {\vec {D}}({{\bf x}},t)\varphi_o + \partial_t  {\vec {D}}({{\bf x}},t)\varphi_o\cr
c\ curl\ {\vec {D}} ({{\bf x}},t) \varphi_o = - \partial_t {\vec B}({{\bf x}},t)\varphi_o.\cr}
\end {cases} 
\end {eqnarray} 
One may indeed remove from the Maxwell-Lorentz system of equations for the electron its density $\rho ({{\bf r}},t)$, which is the function of fields in non-dual physics rather than an independent entity in the empty space paradigm.

By pairing locally energy flows of the radial  fields and the continuous particle densities, one also unifies in (14) gravi-mechanical and electric self-energies $q\varphi_o \Rightarrow Q\varphi_o =  ({\sqrt G} m + ie) c^2/ {\sqrt G} $. Such a double unification (particle with field and gravity with electricity) assumes the application of the complex charge $q\Rightarrow Q = ({\sqrt G} m + ie)$ in the  logarithmic
 interaction  potential (13).  The  many-body system of complex charges $Q_k$ with paired Newton/Coulomb interactions at large inter-particle distances, $R_{lk} >> ( |Q_l|+|Q_k|)/{\varphi}_o$,  possesses real and imaginary energies 
\begin{eqnarray}
 \sum_{l=1}^n\!\!\left ( {Q_l} {\varphi}_o + Q_l\sum^n_{k\neq l} \frac{ (- Q_k)}{2R_{lk} }\right) =
 \cr
\sum_{l=1}^n \!\left [ 
c^2\! \left (\! m_l\! -\! \sum^n_{k\neq l} \frac{ (Gm_lm_k - q_l q_k)}{2R_{lk}c^2 }\!\right)
\!+ \! i  \frac { c^2} {\sqrt G} \left (\! q_l\! -\!  \sum^n_{k\neq l} \frac{ { G}(m_l q_k +  m_k q_l)} {2R_{lk}c^2 }\!\right )
\right]\!\!.
  	\end{eqnarray}

Recall  that the sphere of the microscopic radius $|r_o| =|Q/\varphi_o|$ contains exactly half of the complex electron charge $Q$.  The other half of the elementary astrocharge is distributed over micro, macro, and mega scales in the Universe, which is already known \cite{Asp3} as  the non-local material system. The electron's density scale $|r_o|  \approx e_o{\sqrt G}/ c^2 = 1,38\times 10^{-34}cm $ is even  less than the Planck's length, $l_p \equiv {\sqrt {\hbar c}} ({\sqrt G}/ c^2) = 1,62 \times 10^{-33}cm$, and surely is much smaller of the current limit $10^{-17} cm$ for space inhomogeneity   measurements.  The ratio of classical and quantum scales, $|r_o|^2/l_p^2 \equiv \alpha \approx 1/137$,  can shed some light on the physical origin of the Sommerfeld (fine-structure) constant.
The very small spatial scale $10^{-34}cm$ for main densities of the radial electron explains the formal success of the $\delta$-operator modeling of the nonlocal energy carrier  by the point matter approximation. The formal interaction of such point-like concentrations of energy with  external weak distributions of energy  is conceptually incorrect for extended charges due to their global overlap  in the entire Universe.

One can verify quantitatively that the complex astrocharge distribution, $\rho(r)=Qn(r) = ({\sqrt G} m + ie)n(r)$ with $r_o = Q{\sqrt G}/c^2$,  generates the  potential (13) under the regular integral  rule for classical gravitation and electrodyna\-mics,
\begin{eqnarray} 
\!\int\!\frac{(-Q) n({r}')dv'}{|{\vec r}-{\vec r}'|}
=\!-\!\int_o^\infty\!\int_{-\pi/2}^{\pi/2}\!\int_o^{2\pi}\!\!\frac {d\phi'sin\theta'd\theta'r'^2dr'}{\sqrt {r^2\!+\!r'^2\!-\!2rr'cos\theta'}}\frac{Q r_o}{4\pi r'^2(r'\!\!+\!\!r_o)^2} \cr
= -  \int_o^\infty \frac {dr' Q r_o}{  (r' + r_o)^2} \left(  {{{ {|r' + r|} }\!-\!{ {|r' - r|} }}\over 2rr'}  \right )  
 = - 
 \!\!  \int_r^\infty\!\!  \frac{Qdr'}{ r_o} \left (\frac{1}{r'}\! -\! \frac{1}{r'+r_o}  \right)  \cr
 \! \equiv\! \frac {(-Q)}{r_o} ln \left(1+ \frac {r_o}{r} \right)\!\! =\!\!\int_r^\infty\!\! {{\hat {\bf r}}\rq{}{\vec D}(r')dr'}=W_{qm}(r) .
\end{eqnarray} 
 Notice that $ W_{qm}({r})$ coincides with the work associated with the replacement of a unit probe (positive) charge from the point 
 ${r}$ to  $\infty$ against the radial field  ${\hat {\bf r}}\rq{}{\vec D}(r') = D(r) = - \partial_r W(r)$. The integration over $r'$ within $0 \leq r' \leq r$  vanishes identically in (16) in agreement with the physical meaning of potentials for a probe body.

\section {Maxwell equations are local energy \\ identities in non-dual physics }
  
The exact analytical  solution (10) for Maxwell's equations rigorously requests the $r^{-4}$ radial astroparticle instead of the conventional point particle. This means that 
the empty space paradigm should be replaced by the non-empty space concept for a coherent theory of classical fields. Only distributed elementary energies with smooth  densities should stand behind the non-dual physical reality given in observations of regions with high (former particles, substance) and low (former weak fields) concentrations of energy.  

Like the first pair of Maxwell field identities, $\nabla_\lambda F_{\mu\nu}+
\nabla_\mu F_{\nu\lambda} + \nabla_\nu F_{\lambda\mu} \equiv 0$, the second pair of Maxwell  current equations can also be considered (in the non-empty space paradigm) as identities for balanced momentums  $\rho \varphi_o u^\mu /c$ of (yin-yang) paired energy densities, 
\begin {eqnarray}
\begin {cases} {
[\nabla_\lambda F_{\mu\nu}(x)+
\nabla_\mu F_{\nu\lambda}(x) + \nabla_\nu F_{\lambda\mu}(x)]\varphi_o/4\pi c \equiv 0 \cr
[ \delta^\mu_\lambda - u^\mu(x) u_\lambda (x)] \nabla_\nu F^{\nu\lambda}(x)\varphi_o/4\pi c\equiv 0. \cr }\end {cases}
\end {eqnarray}
Here we used the following identical relations for classical fields:
\begin {eqnarray}
 \begin {cases}
{ 
\nabla_\nu F^{\nu\mu}(x)\equiv  4\pi j^\mu(x)/c \equiv 4\pi \rho(x) u^\mu(x) \cr
u_\mu(x)\nabla_\nu F^{\nu\mu}(x) \equiv 4\pi \rho(x) \cr
\nabla_\nu F^{\nu\mu}(x) \equiv   u^\mu(x) u_\lambda (x)\nabla_\nu F^{\nu\lambda}(x). \cr}
  \end {cases} 
\end  {eqnarray}

  The  energy-momentum balance of paired 4-vector flows in the second identity (17) clams the absence of  curl energy currents along directions normal to the 4 velocity $u^\mu$ of field-energy densities, where $u_{\mu}u^\mu = 1$. These is no much sense to introduce particles at all in the non-dual scheme for field-energy identities  (17). In other words, material  reality is nothing but inhomogeneous energy flows through all spatial points of the Universe.

Once one identified electric currents with field-energy flows in the Maxwell-Lorentz theory for extended electrons, the variation techniques for non-dual material fields should be different from textbook variation procedures of the empty space physics. Indeed, the current Maxwell equations were derived in the dual approach from the summary action of fields and particles varied and fixed independently, like $\delta (A_\mu j^\mu) = j^\mu \delta A_\mu$. In non-dual physics, variations of self-energy of material flows in their potentials $\delta (A_\mu j^\mu)$ are equal to the sum $j^\mu \delta A_\mu + A_\mu \delta j^\mu$, with the double variational result 
\begin {eqnarray} \delta (cA_\mu\nabla_\nu F^{\nu\mu}/4\pi)=(c/4\pi) [\delta A_\mu \nabla_\nu F^{\nu\mu}   + A^\mu\nabla^\nu (\partial_\mu \delta A_\nu - \partial_\nu\delta A_\mu ) ]\cr
\Rightarrow  (c/2\pi) \nabla_\nu F^{\nu\mu}  \delta A_\mu = 2j^\mu  \delta A_\mu. \end {eqnarray}
Therefore, the Lagrangian for non-dual charged energies should posses double electromagnetic densities  $F_{\mu\nu}F^{\mu\nu}$ in the non-empty space action, \begin {equation}
S_{n-em} = - \frac {1}{c} \int d \Omega \left [ \frac {1}{c}A_\mu j^\mu  +  
 \frac {1}{8\pi} F_{\mu\nu} F^{\mu\nu} \right],
\end {equation}
compared to the dual (field + matter) approach.
Then the variations of (20) with respect to four field potentials $A_\mu$ result in the  Maxwell relation  $4\pi j^\mu / c = \nabla_\nu F^{\nu\mu}$ or energy flow identities in  (17) which is the Lagrange equation of motion. It is worth to note, that the non-empty space action (20) vanished on the  Lagrange variation trajectories:
 \begin {eqnarray}
S_{n-em} = - \frac {1}{c} \int d \Omega \left [ \frac {1}{c}A_\mu j^\mu  +  
 \frac {1}{8\pi} F_{\mu\nu} F^{\mu\nu} \right] \cr = -  \int \frac {d \Omega}{8\pi c} \left [ 2 {A_\mu \nabla_\nu F^{\nu\mu} } +  
 (\partial_\mu  A_\nu - \partial_\nu A_\mu ) F^{\mu\nu} \right]
\cr 
 =  -  \int \frac {d \Omega}{4\pi c}  A_\mu \nabla_\nu (F^{\nu\mu} - F^{\mu\nu}) -
 \oint \frac {dS_\mu}{4\pi c}  A_\nu F^{\mu\nu} = 0+0.
 \end {eqnarray}
From here, self interactions of extended charges in electrodynamics balance their electric self-energies like gravitational self-interactions in (9) balance mechanical self-energies $mc^2$ of elementary mass-energy carriers.  This is an expected result that motion along Lagrange trajectories, like in (18), nullifies the action integral. Recall that the conventional current+field electromagnetic action in dual physics with $F_{\mu\nu}F^{\mu\nu}/16 \pi$ instead of $F_{\mu\nu}F^{\mu\nu}/8 \pi$  does not match  such a path nullification in the Lagrange variation formalism.

\section {Discussion}

At first glance the rigid motion of radial fields regarding their
centers of spherical symmetry seems in contradiction
with the light speed limit for signals and retarded physical interactions. Equally, infinite spatial structures of elementary matter seems in contradiction with the speed
limit for information transfer between  bodies. However, any recording/reading of information requires real energy exchanges, while
self-coherent elementary objects maintain their extended
spatial structures without energy losses. 
Indeed, the De Broigle wave oscillations are instantaneous over the entire particle distribution but they can not be transformed into energy signals. 

 Local electric currents and associated mass flows of elementary continuous carriers are proportional to electric energy and mas-energy flows, respectively, of this carrier.  Therefore, the nonlocal Maxwell-Lorentz electron (1)-(3) and (10) tends to replicate (astro)distributions of the  quantum electron in the  De Broglie - Bohm formulation, for instance, of wave mechanics \cite{Boh3}. In particular, De Broglie  periodical oscillations, $\psi = \psi_o exp  i(\omega t - {\bf k}{\bf x})$, of elementary particle\rq{}s matter everywhere in the Universe may be associated in the developed classical approach to the standing wave oscillations  (7) within the continuum electron under the equilibrium distribution of charged densities.

The
Cooper pair, for example, can rigidly keep its self-coherent spatial
distribution over macroscopic regions. No one might say that the
Cooper pair, formed by two electrons, is a point particle. Similarly, each electron may have a rigid spatial
structure with self-coherent phase relations and single-valued distributions of material densities. It doesn't matter that 
normal electrons have different energies and imbalanced
phase rates of De-Broglie oscillations. Like the distributed Cooper pair, each continuous electron possesses  a fine, but finite,
density in all spatial points. Information free elastic reshaping of these elementary densities might occur with internal longitudinal waves without energy losses/gains, 
 while free transverse waves and other dissipative signals (with information and mutual energy exchanges between  overlapping microcosms of elementary energies)
 obey the speed restriction of Special Relativity regarding centers of radial carriers. 

The experimental result \cite {Piz} for the rigid motion of Coulomb\rq{}s
fields regar\-ding  Faraday's \lq{}centers of force\rq{} confirms
the non-empty paradigm for overlap\-ping material bodies. In physical reality, nebular dense fronts  of infinitely extended masses (observed as the planets) are moving within, not outside, the
$r^{-4}$  mass-energy distribution of the radial Sun. Such a global overlap of locally undivided continuous carriers of mass-energy in joint non-empty space explains not only the electromagnetic laboratory experiment \cite {Piz}, but also the gravitational Laplace problem of the Solar system stability.

Again, oscillations of radial energy densities in elementary carriers  may be accompanied by  outward and inward spherical waves (6). Outward waves were considered in the conventional empty space as retarded waves which have already left their point source. Inward spherical waves were called advanced and non-physical waves, because they have not yet met their point source in question. The non-empty space paradigm for radial material sources  provides  equal physical rights to outward and inward waves  in reality of non-local bodies. In this way, intercepted longitudinal waves along advanced directions toward moving centers of cosmic bodies \cite{Koz} may be considered as a justification of the non-empty space in reality.  Similarly, inward radial signals might be expected under acceleration of electric charges when their radial structures attain external energy.  It appears that an interesting development of experiments \cite {Piz} might be their reiteration with proton beams at the Large  Hadron Collider. Already available  energies of beams might not only confirm the rigid motion of a Coulomb radial field, found for electron beams, but might also reveal inward waves toward future positions of non-equilibrium protons under acceleration.  In general,  wave signals  along a computed advanced axis toward moving energies in gravitation and electrodynamics  may ultimately justify the non-empty space paradigm for nonlocal world reality.



\end{document}